# Capturing collective conflict dynamics with sparse social circuits

EDWARD D LEE, BRYAN C DANIELS, DAVID C KRAKAUER & JESSICA C FLACK, Center for Complexity & Collective Computation, Wisconsin Institute for Discovery

## 1. INTRODUCTION

A central problem in the study of collective behavior is how functionally significant macroscopic properties arise out of individual or microscopic interactions. The most common approach to studying the relationship between the micro and macro in biological systems is dynamical systems and pattern formation (for examples, see Sumpter, 2006; Ball, 2009; Couzin, 2009; Payne et al., 2013). A complementary approach is to treat the micro to macro mapping explicitly as a computation.

A macroscopic property can be said to be an output of a computation if it can take on values that have functional consequences at the group or component level, is the result of a distributed and coordinated sequence of component interactions under the operation of a strategy set, and is a stable output of input values that converges (terminates) in biologically relevant time [Flack and Krakauer 2011][Flack 2014].

The input to the computation is the set of elements implementing the rules or strategies. The input plus the strategies constitute the system's microscopic behavior. My collaborators and I have developed novel computational techniques, called Inductive Game Theory [DeDeo et al. 2010][Flack and Krakauer 2011][Lee et al. 2014], to extract strategic decision-making rules from correlations observed in the time series data and reconstruct the microscopic behavior.

In biological systems, there are typically multiple components interacting. Hence the computation of the macroscopic output is inherently collective, meaning we must examine how different configurations of strategies affect the macroscopic output [Flack 2014]. We describe the space of microscopic configurations using Markovian, probabilistic, "social" circuits [DeDeo et al. 2010][Flack and Krakauer 2011][Lee et al. 2014].

Here, we briefly illustrate this approach using time series data collected on conflict dynamics from an animal society model system—a large, captive group of captive pigtailed macaques (*Macaca nemestrina*) (n = 47). We start with very simple time series data on which individuals were present in a conflict. This gives us a time series of binary fight participation vectors. Given this input of individual identities, our output is the distribution of fight sizes, which has been shown to have functional consequences for individuals in the study system [DeDeo et al. 2010]. We ask whether individuals decide to join fights based on participation in the last fight (say between their allies and adversaries), what are their decision-making strategies, and how do these strategies collectively produce the distribution of fight sizes?

## 2. STRATEGY EXTRACTION AND CIRCUIT CONSTRUCTION

We begin by defining a space of strategies. Since we are interested in how individuals join fights and have limited a priori our input data to identity of fight participants, we can represent this space of strategies as $c(n, m)$, where $m$-tuplet of individuals in the fight at time $t$ is determined by the behavior of an $n$-tuplet of individuals in the fight at time $t-1$ [DeDeo et al. 2010]. We restrict the $C(n, m)$ space we explore to $C(1,1)$, $C(2,1)$, and $C(1,2)$ the simplest strategies in the space as cognitive and behavioral data indicate that the macaques are capable of using these strategies.

Choosing $C(n,m)$ specifies the identity of the vertices between which can allow for strategic connections. The directed edges correspond to strategic information from $m$ to $n$ and we compute these as

$$\Delta P(i_{t-1} \to j_t) = \frac{N(j_t|i_{t-1}) - N_{\text{null}}(j_t|i_{t-1})}{N(i)} \quad (1)$$

$N(i)$ is the total number of events in which the $n$-tuplet $i$ participated at all times $t-1$, and $N(j_t|i_{t-1})$ is the number of events in which the $m$-tuplet $j$ participated at $t$ given $i$ in the previous event. $N_{\text{null}}$ is the average calculated over time permutations of the fights while maintaining the participation integrity of each event (*e.g.* we only shuffle the order of events; we do not shuffle participation).

The $\Delta P$ corresponds to a measurement of how often $j$ follows $i$ as a deviation from shuffle time series. Positive (negative) values indicate that $j$ is more (less) likely to follow $i$ than by chance. Since $\Delta P$ is normalized to $-1$ to 1, we interpret these values probabilistically where negative values correspond to a suggestion of strategic avoidance in the following fight.

We use the list of $\Delta P$ found for each $C(n,m)$ strategy class to construct a social circuit for each strategy class. By modulating how we account for these causal relationships, we end up with multiple alternative circuits, or a family of circuits for each $C(n,m)$ strategy class. Each circuit in the family is a model of the micro-macro relationship and so serves as a hypothesis for how strategies combine over nodes (inputs) to produce to the target output [Lee et al. 2014]. We test the circuits against each other in simulation to determine which can best recover the actual measured macroscopic behavior of our system and other fine grained statistics.

We find that (1) a variant of the $C(2,1)$ circuit, with the measured $\Delta P$, captures some important features of the microscopic behavior of the system, (2) but that the distribution of fight sizes is degenerate over configurations of the $C(2,1)$ circuit parameter space. Although our analyses support the conclusion that the distribution of fight sizes results from the summed contributions of the $\Delta P$ in the $C(2,1)$ circuit, we note that the degeneracy of the distribution of fight sizes for $C(2,1)$ parameter space coupled to the observation that the $C(2,1)$ circuit is extremely complex with many precise, tiny parameters, suggests that the exact measured values of $\Delta P$ may not matter. A natural next step is to systematically reduce the dimensionality of the circuit. Our goal is two-fold: to find the a parsimonious description of the generative process that is also consistent with what we know about the information the monkeys use to make decisions—a *cognitive effective theory*.

## 3. SPARSE SOCIAL CIRCUIT CONSTRUCTION

Our approach is to compress the time series data to build a more compact circuit. Using a technique from neuroscience called sparse coding [Olshausen and Field 1997], we extract the individuals and subgroups that are regular and predictable conflict participants [Daniels et al. 2012]. We show an example from this process in Figure 3. Between these sparse groups, we search for $\Delta P$ strategies as specified in Eq 1, a space we denote as $C(s_i, s_j)$, where the sizes of groups $1 \leq |s_i|, |s_i| \leq 5$. Consistent with previous results [DeDeo et al. 2010], we find more significant $C(2,1)$ and less significant $C(1,1)$ strategies than would be expected from a null permutation of the data. On the other hand, we also find evidence for other strategy classes beyond the three explored in previous work.

Since we expect that important social groups in the data set constitute relevant features for strategy formation, we also allow for strategies of the form $C(s_i, 1)$, that is information that individuals use to form strategic decisions based on previous participation of a sparse group. Even with this addition, we have $\sim$12 sparse groups and so the number of strategic edges that can exist on this graph are several orders smaller than the bipartite $C(2,1)$, a lesser cognitive burden on individuals who might use a similar model to predict conflict behavior.

Compared to other models that have been proposed for this system, the $C(s_i, s_j\&1)$ sparse circuit constitutes a comparatively reasonable fit despite being far more cognitively and computationally parsimonious in the number of possible strategies.

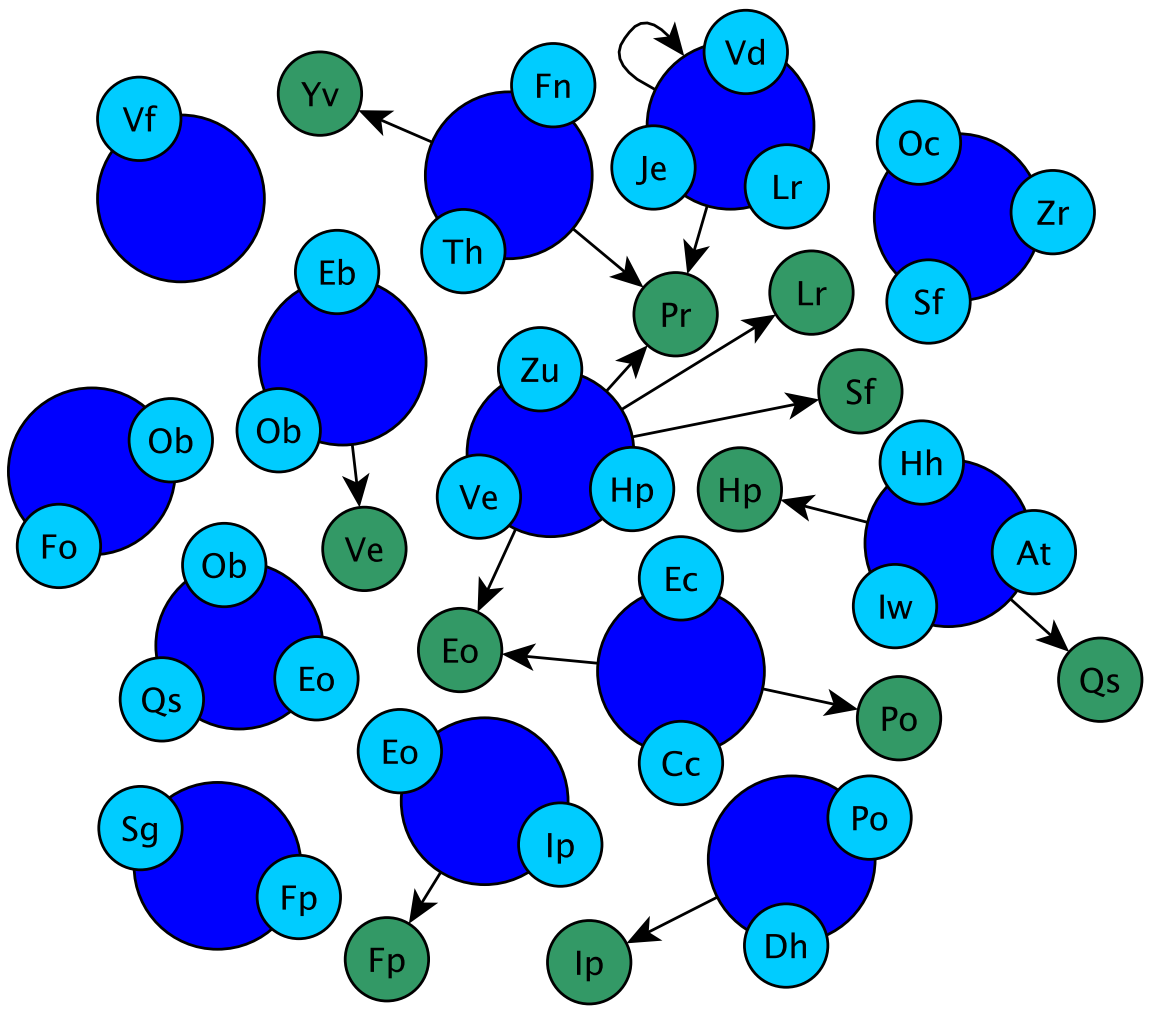

Fig. 1. Example of sparse basis circuit obtained by sparse coding. Individuals in light blue circles are part of groups in dark blue. Individuals not in basis groups are green. For example, individual Vf is in a basis group alone. Eo and Ob appear in multiple different basis groups in this solution. We only show the top 15 largest $\Delta P$ significant with $p \leq 0.05$; these are all positive. This circuit is simpler than the $C(2,1)$ network in so far as it has many fewer nodes and these nodes are composed of the regular and predictable individuals.

## 4. CONCLUSION

We have proposed an approach for reconstructing the microscopic, strategic behavior of components and individuals in social systems. This approach involves extracting component decision-making strategies from time series data on competitive interactions and constructing circuits that specify how these strategies combine to produce functionally important macroscopic properties that describe social structure.

With the exception of a few life history variables, it is not yet known what the important macroscopic properties are in biological and social systems. As we refine our understanding of the micro to macro mapping through construction of "full" circuits capturing the detailed microscopic behavior and prune these circuits to construct cognitive effective theories for the macroscopic output, we gain insight into the what properties of the time series data constitute the "right" input and hence the building blocks of our system. By investigating whether our most accurate, empirically justified circuits can also account for other potentially important macroscopic properties, we can begin to establish which macroscopic properties might be fundamental and what their relation is to one another, thereby moving towards a thermodynamics of biological collectives [Flack 2014].

## REFERENCES

Jeanne Altmann. 1974. Observational study of behavior: sampling methods. *Behaviour* 49 (1974), 227–267.

Nihat Ay, Jessica C Flack, and David C Krakauer. 2007. Robustness and complexity co-constructed in multimodal signaling networks. *Philosophical Transactions of the Royal Society* 1479 (2007), 441–447.

Julian O Caldecott. 1986. An ecological and behavioural study of the pig-tailed macaque. S. Karger.

Bryan C Daniels, David C Krakauer, and Jessica C Flack. 2012. Sparse code of conflict in a primate society. *Proceedings of the National Academy of Sciences* 109, 35 (2012), 14259–14264.

Marjolijn Das. 2000. Conflict management via third parties. In *Natural Conflict Resolution*, F. Aureli and F.B.M. de Waal (Eds.). University of California, Berkeley, 263–280.

Frans B M De Waal and Peter L Tyack (Eds.). 2011. *Animal social complexity*. Harvard University Press, Cambridge, MA.

Simon DeDeo, David C Krakauer, and Jessica C Flack. 2010. Inductive game theory and the dynamics of animal conflict. *PLoS Computational Biology* q-bio.PE, 5 (June 2010), e1000782.

Simon DeDeo, David C Krakauer, and Jessica C Flack. 2011. Evidence of strategic periodicities in collective conflict dynamics. *Journal of The Royal Society Interface* 8, 62 (July 2011), 1260–1273.

Jessica C Flack. 2012. A future for social evolution (comment on Gintis–the Evolution of Cooperation). *Cliodynamics* 3 (2012), 172–190.

Jessica C Flack. 2014. Life's information hierarchy. *Santa Fe Bulletin* (2014).

Jessica C. Flack and Frans B M de Waal. 2007. Context modulates signal meaning in primate communication. *Proc. Natl. Acad. Sci* 104 (2007), 1581–1586.

Jessica C Flack, Frans B M de Waal, and David C Krakauer. 2005. Social structure, robustness, and policing cost in a cognitively sophisticated species. *The American Naturalist* 165, 5 (2005), E126–39.

Jessica C Flack, Michelle Girvan, Frans B M . De Waal, and David C Krakauer. 2006. Policing stabilizes construction of social niches in primates. *Nature* (2006).

Jessica C Flack and David C Krakauer. 2006. Encoding power in communication networks. *The American Naturalist* 168, 3 (Sept. 2006), E87–102.

Jessica C Flack and David C Krakauer. 2011. Challenges for Empirical Complexity Measures: A Perspective from Social Dynamics and Collective Social Computation. *Chaos: An Interdisciplinary Journal of Nonlinear Science* 21, 3 (2011), 037108.

Jessica C Flack, David C Krakauer, and Frans B M de Waal. 2005. Robustness mechanisms in primate societies: a perturbation study. *Proceedings of the Royal Society, Series B Biological sciences* 272, 1568 (2005), 1091–1099.

Achternaam H Harcourt and Frans B M de Waal (Eds.). 1992. *Coalitions and alliances in humans and other animals*. Oxford University Press, Oxford.

David C Krakauer, Jessica C Flack, Simon DeDeo, and Doyne Farmer. 2010. Intelligent Data Analysis of Intelligent Systems. *IDA 2010, LNCS 6065* (2010), 8–17.

David C Krakauer, K Page, and Jessica C Flack. 2011. The Immuno-Dynamics of Conflict Intervention in Social Systems. *PloS One* 6 (2011), e22709.

Edward D Lee, Bryan C Daniels, David C Krakauer, and Jessica C Flack. 2014. Capturing collective conflict dynamics with sparse social circuits. (2014).

Martin A Nowak. 2006. Five rules for the evolution of cooperation. *Science* 314 (2006), 1560–63.

Bruno A Olshausen and David J Field. 1997. Sparse coding with an overcomplete basis set: A strategy employed by V1? *Vision Research* 37, 23 (1997), 3311–3325.

John M Smith. 1976. Evolution and the theory of games. *American Scientist* 61 (1976), 41–45.

Gabor J Szekely and Maria L Rizzo. 2009. Brownian distance covariance. *The Annals of Applied Statistics* 3, 4 (2009), 1236–1265. DOI:http://dx.doi.org/10.1214/09-AOAS312

Bernard Thierry. 2000. Conflict management patterns across macaque species. In *Natural Conflict Resolution*, F. Aureli and F.B.M. de Waal (Eds.). University of California Press, Berkeley, 106–128.

John von Neumann and Oskar Morgenstern. 1944. *Theory of Games and Economic Behavior*. Princeton University Press, Princeton, NJ.